\documentstyle[11pt,newpasp,twoside,epsf]{article}
\newcommand{\lsi}    {LS~I+61$^{\circ}$303}

\def\edcomment#1{\iffalse\marginpar{\raggedright\sl#1\/}\else\relax\fi}
\reversemarginpar

\begin{document}
\title{   Confirmation of a moving component in
   the H$\alpha$ emission line of LS~I+61$^{0}$303 }
 \author{ Radoslav K. Zamanov}
\affil{ National Astronomical Observatory Rozhen,
        4700 Smoljan, Bulgaria }
\author{ Josep Mart\'{\i} }
\affil{
Dep. F\'{\i}sica, Univ. de Ja\'en,
Virgen de la Cabeza 2, 
23071 Ja\'en, Spain}

\begin{abstract}
We report our attempts to detect and confirm a narrow moving component in
the $H\alpha$ emission line of \lsi. The existence of this spectral feature
was already suspected in the past. As a result, we find that this
component does exist and that its radial velocity varies in agreement
with the radio period of the system. We interpret it tentatively as
due to a denser region, or bulge, orbiting near the outer edge of the
H$\alpha$ emitting disk.
\end{abstract}


\section{Introduction}

LS~I+61$^{0}$303 (V615 Cas, GT~0236+610)  is a radio emitting Be/X-ray 
binary, with its primary being
a rapidly rotating B0V star with an equatorial disk. The secondary is
a compact object, most probably a neutron star, orbiting in an eccentric
orbit. The most spectacular phenomena associated with this object are
its periodic non-thermal radio outbursts, repeating every $\sim 26.5$ d.
This interval is supposed to be the orbital period of the binary.

The H$\alpha$ of \lsi\ is observed as a double peaked emission
line of variable profile intensity which is not peculiar with respect
to those observed in other Be stars
High resolution H$\alpha$ observations of this object are discussed
in few papers (Paredes et al. 1994, Zamanov et al. 1999).
In these papers, variability of the H$\alpha$ emission line was established over
time scales of days and
two possible interpretations were proposed to account for it,
namely: (i) an influence of the compact object onto the Be star
disk; (ii) an unresolved emission line component due to gas
within the system or ejected from it.

The present paper discusses $H\alpha$ observations of \lsi\
with the aim of testing the reality
of proposed unresolved component. 

\section{ Observations, data reduction and results}

The observations were obtained with the Coud\'e
spectrograph of the 2 m telescope at the Bulgarian NAO ``Rozhen''. 
They cover about 110 \AA\ (0.2 \AA\ pixel$^{-1}$) and
have a signal-to-noise ratio in the range 30-50.
In order to detect a possible narrow or unresolved
component we have applied the following procedure:

\begin{enumerate}

\item   All spectra were transformed to an uniform heliocentric 
wavelength set.

\item  The spectra were divided into 10 groups. Every group 
consists of 3-4 spectra, obtained in consecutive or close nights. 
For each group the ``minimum spectrum" was generated by taking 
the minimum pixel intensity
from all the spectra of the group over the uniform wavelength set.

\item  The ``minimum spectrum"  was subtracted from all spectra in the
group. The residuals were smoothed with a moving average over
10-15 points (2 to 3 \AA). An example  is shown in Fig.~1.

\end{enumerate}

The applied procedure assumes explicitly that, if moving component exists, 
it must appear as a pure emission line (not absorption).

 \begin{figure}
 \epsfysize=6.0cm
\vspace*{60mm}
\begin{minipage}{60mm}
\includegraphics{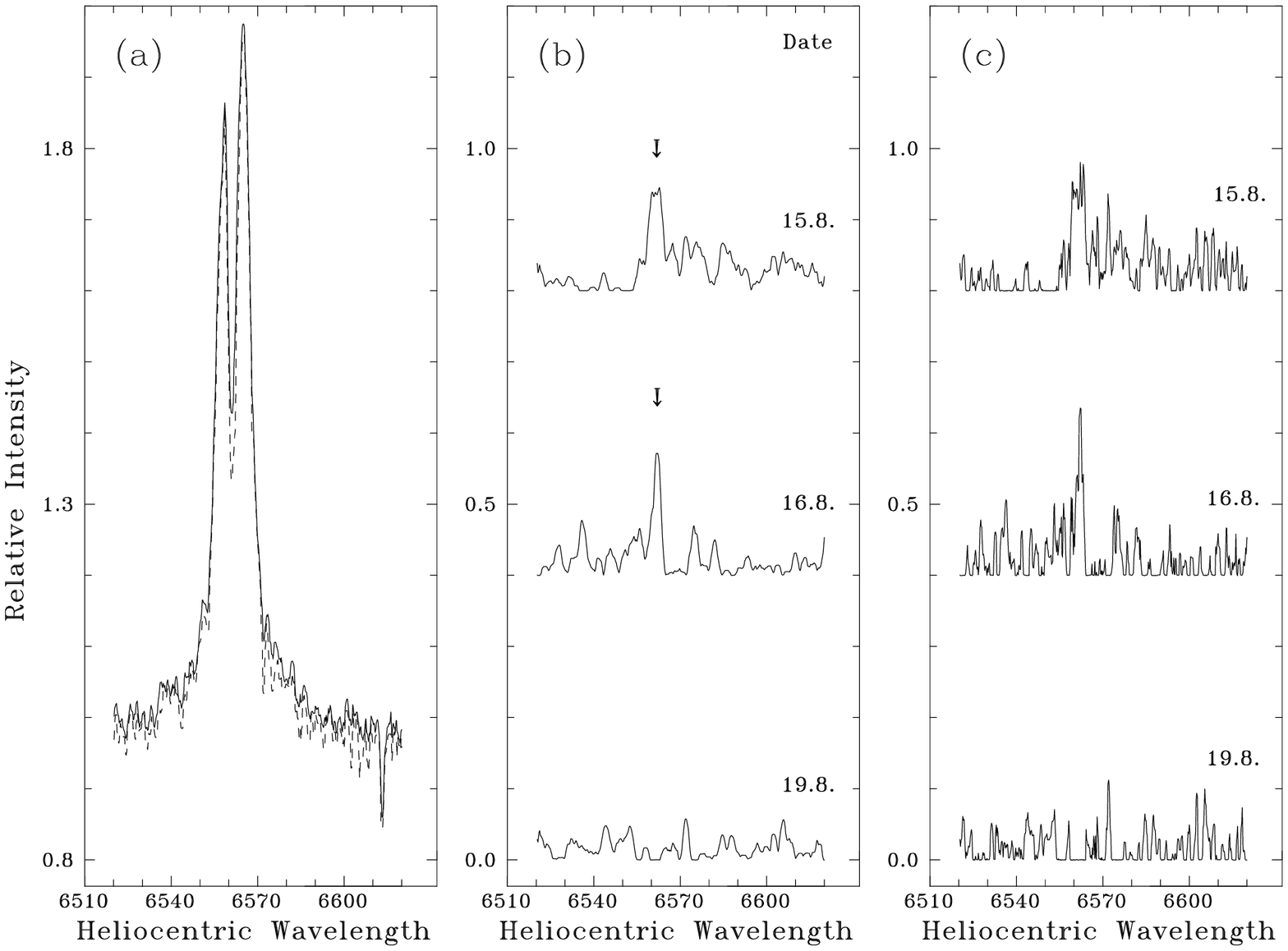}
\end{minipage}
     \caption
          {
   {\bf (a)} An example of average spectrum (solid line) and
   the ``minimum spectrum" (dashed line) obtained using the spectra 
   from 15, 16 and 19 August 1997.
   {\bf (b)} The smoothed residuals for the same group. The arrows
   indicate the positions of the moving component.
      {\bf (c)} The unsmoothed residuals for the same spectra.
The plots are offset by adding arbitrary constants
identical for panels (b) and (c).
         }
    \end{figure}

It is visible in Fig. 1 that a moving narrow component does exists. This feature
clearly shows up in most residual plots after the ``minimum spectrum" is subtracted.
The typical FWHM of this moving component is  4-8 \AA,
with a normalized intensity about 0.1-0.2. The radial velocity of the component
was determined by employing a Gaussian fit, with a 
typical error of $\pm 50$ km s$^{-1}$.
We plotted our measurements in the left panel of Fig. 2 folded on the
radio phase of \lsi. The latest ephemeris by Gregory et al. (1999) was used for this purpose, with
the phase origin set at JD 2443366.775 and a 26.4917 d radio period being adopted.
A clear trend of the component radial velocity emerges in this figure. Indeed,
a modulation with the same 26.5 d radio period and a $\pm250$ km s$^{-1}$ amplitude
seems to be present in the data.
Such a strong radial velocity dependence gives us confidence that we are
detecting a real variability, and that this is not likely to be an
artifact of the applied procedure.
It deserves to be noted here that the moving component was already directly 
visible in the high resolution spectra of Paredes et al. (1990), 
and without any subtraction procedure being applied. The radial velocity
measurements of these authors have been incorporated in Fig. 2 and
they are in excellent agreement with our observations.

By fitting to the data points a least squares cosine function 
of the form:
$\;\; V_{\rm r}(\varphi)=A+B\cos{[2\pi(\varphi-\varphi_0)]}, \;$
we obtain the following best fit parameters: $A=-21\pm14$ km s$^{-1}$,
$B=186\pm20$ km s$^{-1}$, and a phase shift $\varphi_0=0.68\pm0.02$
($\varphi$ is the radio phase). As it can be expected, the parameter
$A$ is near to the system velocity $-55$ km s$^{-1}$ derived
by Hutchings \& Crampton (1981). The best fit cosine fit is plotted
in Fig. 2 as a solid line.
 \begin{figure}    
 \epsfysize=4.0cm
\vspace*{50mm}
\begin{minipage}{50mm}
\includegraphics{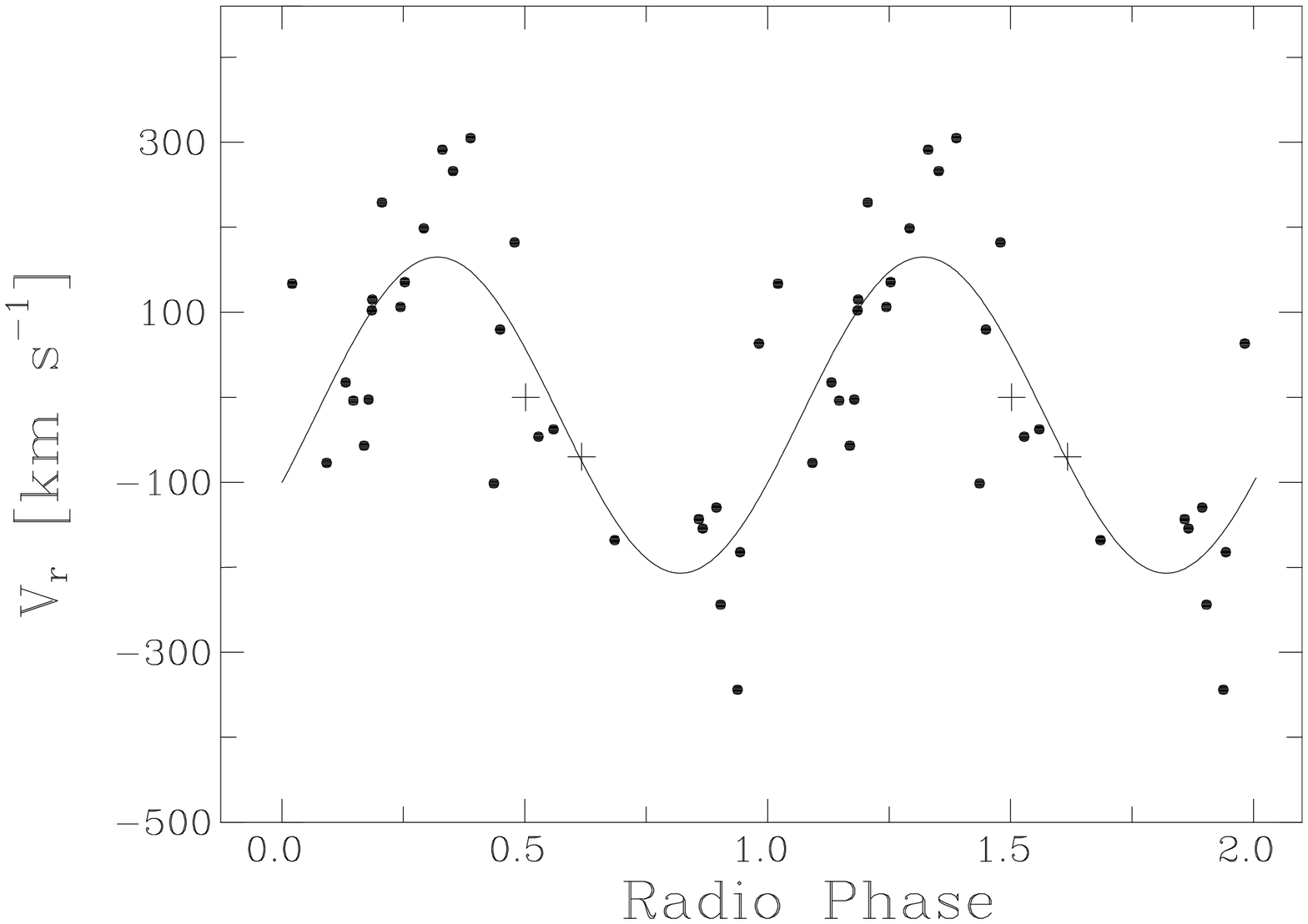}
\includegraphics{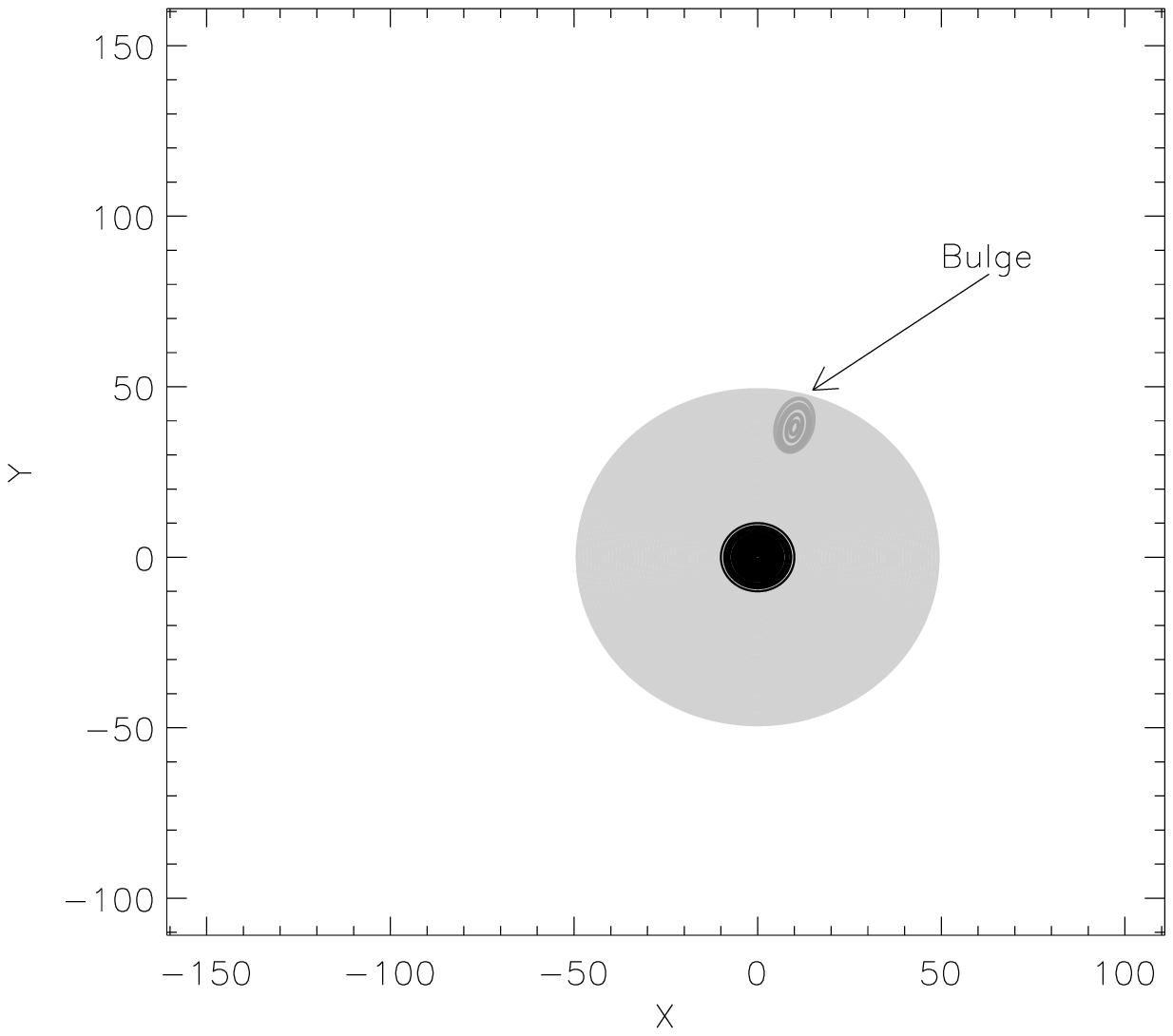}
\end{minipage}
     \caption[] {
{\bf Left panel:}  Phase plot of the radial velocities of the moving  
   component as a function of radio phase. 
   The solid line is the best cosine fit.
   The dots represent our data, the crosses - from Paredes et al. (1990). 
{\bf Right panel:}  Schematic representation of the Be star, 
its H$\alpha$-emitting disk, and the bulge located near to the outer edge. 
The axes are in units of $R_{\odot}$.  }
    \end{figure}

\section{ Discussion}

What is the origin of the moving component?
Among possible interpretations, one could think of a possible relationship with
the compact companion in the \lsi\ system, i.e., due to matter captured 
by the neutron star from the Be star wind - the material around the neutron star,
or an accretion disk, or an accretion wake.

The orbital parameters of \lsi\ are not well constrained so far.
The proposed orbits by Hutchings \& Crampton (1981) and Mart\'{\i}
\& Paredes (1995) are quite different.
In spite of this uncertainty, the observed variations up to $\pm 250$ km s$^{-1}$
are significantly larger than those expected from the orbital
motion of the neutron star, as well as of the Be star, in all
proposed orbital solutions. Consequently a direct relationship, in the sense that
the moving component is reflecting the motion of the neutron star,
does not seem to apply unless the orbit is extremely eccentric,
and alternative interpretations need to be considered.

The spectrally resolved interferometry of $\zeta$ Tauri
by Vakili et al. (1998) has shown a bulge of emission in the disk of this Be star.
These authors have interpreted their interferometric and H$\alpha$ observations
assuming that such a bulge follows a circular orbit in the equatorial
disk plane, at a distance of about 7 stellar radii from the central star.
In this context,  it is thus  conceivable to us that the moving H$\alpha$ component in \lsi\
is also due to a bulge region existing in the circumstellar disk.

Under the assumption that this is the case, let us find the position of the bulge.
Hereafter, we will adopt the following parameters for the Be star in \lsi:
$M_*=10\;M_\odot$, 
$R_*=10\:R_\odot$, and $v\sin{i}=360$ km s$^{-1}$
(Hutchings \& Crampton 1981).
Assuming that the star rotates at 90\% of its critical velocity
(Sletteback et al. 1992),
we derive $\sin{i}=0.92$ and $i \simeq 67^{\circ}$.
Consequently, from the term $B$ in Eq.~1
and the above assumptions we obtain that the bulge in \lsi\
rotates at a velocity $V_{\rm bulge}=B/\sin{i}=202$ km s$^{-1}$. This
corresponds to a distance of about $\sim 4.5$ $R_*$, provided
that the bulge is in a Keplerian orbit around the star.

From the relationship between the separation of the H$\alpha$ peaks,
$\Delta V_{\rm peak}$, and the outer radius $R_{\rm out}$ of a Keplerian emitting disk:
$
 R_{\rm out}=R_*(2 v \sin{i}\:/\:\Delta V_{\rm peak})^2,
$
we find that $R_{\rm out}=4.0$-6.6 $R_*$
(using $\Delta V_{\rm peak} = 300$-360 km s$^{-1}$, Zamanov et al. 1999).
This result compares satisfactorily to the bulge distance found above 
and implies that the bulge must be located near to the outer edge 
of the $H\alpha$ emitting disk.
A schematic picture of the $H\alpha$ disk and the position of the bulge
is represented in the right panel of Fig. 2.

As mentioned before, the orbital elements of the Be X-ray binary
\lsi\ are not well determined. 
From the 26.5 d period present
in the $V_{\rm r}$ data, one could wonder if the position of
the bulge should be in phase with the motion of the neutron star. However,
this does not occur in any of the proposed orbital solutions.
Whether at the end there is a relationship between the bulge position and
that of the neutron star remains to be determined by a careful revision
of orbital solutions in the future. \\

 {\it Acknowledgements:} RZ acknowledges support by IAU Grant 
and by Bulgarian NSF (MUF-05/96).
JM is supported by DGICYT (PB97-0903) and by Junta de Andaluc\'{\i}a (Spain).

\begin{table}
\caption[]{Radial velocities of the moving component }
\begin{center}
\begin{tabular}{|c|c|c|c|c|c|}
\hline
JD2400000+ & $V_r [km/s]$ & Group & JD2400000+ & $V_r [km/s]$ & Group  \\
\hline
 49292.31  &    266 & 1  &      50293.55  &    -77 & 7  \\
 49293.27  &    305 & 1  &      50294.57  &     17 & 7  \\
 49294.53  &   -101 & 1  &      50298.58  &     -- & 7  \\
 49354.21  &   -168 & 2  &      50320.58  &     -- & 8  \\
 49356.26  &     -- & 2  &      50321.53  &     -3 & 8  \\
 49374.48  &     79 & 3  &      50322.37  &     -2 & 8  \\
 49375.25  &    182 & 3  &      50322.53  &    102 & 8  \\
 49385.33  &   -143 & 3  &      50322.56  &    114 & 8  \\
 49386.26  &   -129 & 3  &      50676.52  &     31 & 9  \\
 49528.45  &    135 & 4  &      50677.36  &    -26 & 9  \\
 49529.47  &    198 & 4  &      50680.56  &    187 & 9  \\
 49530.48  &    291 & 4  &      50685.49  &   -222 & 9  \\
 49531.42  &     -- & 4  &      50686.47  &   -260 & 9  \\
 50181.29  &     -- & 5  &      50687.55  &   -197 & 9  \\
 50182.26  &     -- & 5  &      50688.59  &    -76 & 9  \\
 50183.35  &   -344 & 5  &      50689.60  &     77 & 9  \\
 50242.53  &    -57 & 6  &      50703.38  &   -293 & 10 \\
 50243.52  &    229 & 6  &      50704.56  &    326 & 10 \\
 50244.52  &    106 & 6  &      50705.54  &    216 & 10 \\
\hline
\end{tabular}
\end{center}
\end{table}
\end{document}